\documentstyle[twoside,fleqn,espcrc2,epsf]{article}

\newcommand{\AmS}{{\protect\the\textfont2
  A\kern-.1667em\lower.5ex\hbox{M}\kern-.125emS}}

\title{
\hfill\begin{minipage}{0pt}\scriptsize\vspace*{-1.5cm} \begin{tabbing}
\hspace*{\fill} UTHEP-387\\ 
\hspace*{\fill} UTCCP-P-47 
\end{tabbing} 
\end{minipage}\\[-8pt]Two-dimensional Gross-Neveu Model with Wilson Fermion Action \\
at Finite Temperature and Density\thanks{Talk presented by J. Noaki.} }

\author{T. Izubuchi$^{\rm a}$, J. Noaki $^{\rm a}$and A. Ukawa
\address{Institute of Physics, University of Tsukuba, Tsukuba, 
Ibaraki 305-8571, Japan \\ 
${}^{\rm b}$Center for Computational Physics, University
        of Tsukuba, Tsukuba, Ibaraki 305-8577, Japan}$^{\rm ,b}$}
        
\begin{document}

\begin{abstract}
We analytically investigate the 2-dimensional Gross-Neveu model at 
finite temperature and density using Wilson fermion action.  
The relation between the phase structure on the lattice and that in the 
continuum is clarified.
\end{abstract}

\maketitle

\section{INTRODUCTION}

The 2-dimensional Gross-Neveu model\cite{Gross-Neveu} has often been 
used to examine theoretical issues of QCD.  
This stems from the fact that the model not only shares the property of  
asymptotic freedom and spontaneous breaking of chiral symmetry with
QCD, but also can be solved analytically in $1/N$ expansion.  
In this article we present a summary of an yet another 
study\cite{INU} of this category. 
The problem addressed is how the phase diagram of the continuum theory 
at finite temperature and density\cite{Wolff} 
emerges from that of the lattice theory 
formulated with the Wilson fermion action.  In particular we 
explore how this limit is achieved in the presence of the 
parity-broken phase due to the Wilson action\cite{Aoki,AUU}.   
An additional interest is to gain an understanding on how the tricritical 
point of the continuum model, separating a first- and second-order 
chiral transition on the $(T,\mu)$ plane, emerges from the lattice point 
of view. 

\section{LATTICE GROSS-NEVEU MODEL}
Our lattice model is given by the Lagrangian, 
\begin{eqnarray}
{\cal L}&=&\frac{1}{2a}\bar{\psi}_n\gamma_1(\psi_{n+\hat{1}}
-\psi_{n-\hat{1}}) \nonumber \\
&+ &\frac{1}{2a}\bar{\psi}_n\gamma_2(e^{-\mu a}\psi_{n+\hat{2}}
-e^{\mu a}\psi_{n-\hat{2}})+\delta m \bar{\psi}_n \psi_n \nonumber \\
&-&\frac{1}{2N}
[g_\sigma^2 (\bar{\psi}_n \psi_n)^2+g_\pi^2 (\bar{\psi}_n i\gamma_5 \psi_n)^2]\nonumber\\
&+& \frac{a}{2}\bar{\psi}_n \nabla_\mu^2\,\psi_n
\end{eqnarray}
where $\psi$ is an $N$-component fermion field, 
and the chemical potential $\mu$ is 
introduced in the standard manner. In order to absorb 
the effect of the Wilson term $\bar{\psi}_n \nabla_\mu^2\,\psi_n$ 
which breaks chiral symmetry, a mass 
counter term $\delta m$ is introduced\cite{Eguchi-Nakayama},  and
the two interaction terms are assigned an independent coupling constants
$g_\sigma^2$ and $g_\pi^2$\cite{Aoki-Higashijima}. 

In the large $N$ limit, one 
can analytically calculate the effective potential $V$ as a function of 
$\sigma\equiv -g_\sigma^2\bar{\psi}\psi/N+\delta m$ and 
$\pi\equiv-g_\pi^2\bar{\psi}i\gamma_5\psi/N$. 
Making an expansion of the effective potential in terms of $a$, one finds 
that the continuum limit with chiral symmetry requires the 
following tuning relations of the coupling parameters\cite{Aoki-Higashijima}:
\begin{eqnarray}
\frac{1}{g_\pi^2} &=&\frac{1}{g_\sigma^2}+4C+{\cal O}(a),\label{tune1}\\
\delta m a&=& -2C\,g_\sigma^2+{\cal O}(a^2),\label{tune2}
\end{eqnarray}
where $C=2\sqrt{3}/27+1/(12\pi)$. 
Therefore we have to study the lattice phase
structure in the 3-dimensional space spanned by 
$(g_\sigma^2,g_\pi^2,\delta m a)$.

\begin{figure*}[htb]
\epsfxsize=1.0\textwidth
\epsfbox{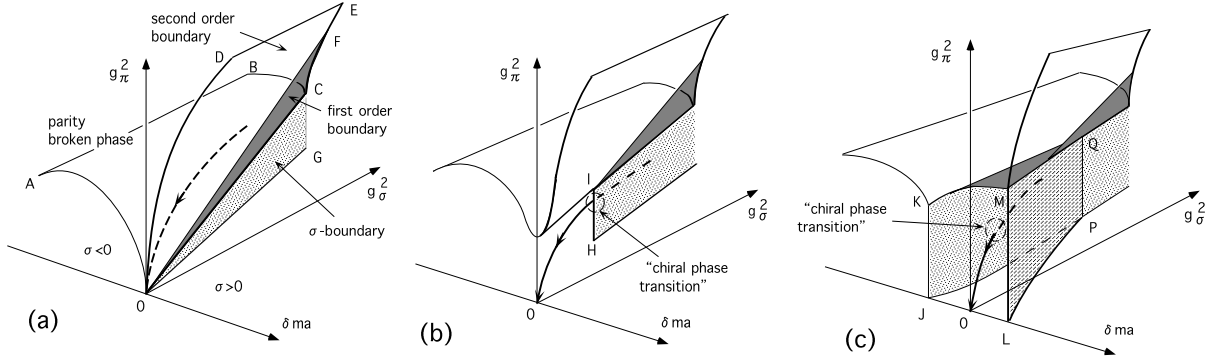} 
\vspace{-0.5cm}
\caption{Lattice phase structure at (a) $T=\mu=0$, (b) $T\neq 0$ and
 $\mu=0$ with finite $N_T$, (c) $T=0$ and $\mu\neq 0$ with finite $\mu a$. 
The (dashed) curve with arrow shows approach to the continuum 
limit (point O).}
\vspace{-0.4cm}
\end{figure*}
 
\section{LATTICE PHASE STRUCTURES AND THE CONTINUUM LIMIT}

We analyze the phase structure in terms of the effective potential 
$V(\sigma, \pi)$ and the saddle-point equations 
$\partial V/\partial\sigma =0$, 
$\partial V/\partial\pi=0$ calculated to leading order in $1/N$. 
The latter equations exhibit both parity-broken 
solution $\pi\ne 0$ and symmetric solution $\pi=0$.  One may expect that the 
phase boundary between the two solutions is determined by taking the limit 
$\pi\to 0$ from the parity-broken phase, in which case the transition would be 
of second order.  The structure of the phase boundary is actually much more 
complex, some part of which turns out to be of first order after evaluation 
of the effective potential to locate the true minimum, as described 
below\cite{INU}.

\subsection{$T=0$ and $\mu=0$}

The phase structure for $T=0$ and $\mu=0$ is schematically shown in Fig.~1(a). 
There are two regions separated 
by the boundary surfaces OABC and ODEC which join along the line OC. 
In the upper region, parity is spontaneously broken, while the system is 
parity-symmetric in the lower region.   
The parity-breaking phase transition along these 
surfaces  is of second order except on the gray part OFC on which it is of 
first order.  This gray part shrinks toward the continuum limit (point O), 
hence has no effect on the continuum physics.  

The parity-symmetric phase in the lower region is divided into two
phases by the hatched surface OCG which is vertical to 
the $(g_\sigma^2, \delta ma)$ plane. The value of $\sigma$ jumps from 
$\sigma<0$ to $\sigma>0$ across this surface, hence the transition is 
of first order.  Hereafter, we call it as $\sigma$-boundary. 

The continuum limit is taken along the curve specified by 
(\ref{tune1})-(\ref{tune2}).  If we take the leading term ({\it i.e.,} 
ignore ${\cal O}(a)$ and ${\cal O}(a^2)$ terms), 
we obtain the the dashed curve in Fig.~1(a) which runs 
just below the surface OABC.  The ${\cal O}(a)\times{\cal O}(a^2)$ 
ambiguity, however, allows us to shift the dashed curve to the 
right so that it runs below the surface ODEC.  In either case, physical 
quantities such as the value of $\sigma$ converge to the continuum value. 
In particular, one can explicitly check that 
the first-order transition along the gray part does not 
cause problems in taking the continuum limit\cite{INU}. 

\subsection{$T\neq 0$ and $\mu=0$}

When we consider the system for a finite temporal lattice size, $N_T$, 
corresponding to a finite temperature $T=1/(N_Ta)$, 
the parity-breaking phase boundary shifts in the positive $g_\pi^2$ direction. 
At the same time the $\sigma$-boundary recedes away from the point O 
so that a gap opens, as illustrated in Fig.~1(b). 

When we approach the continuum limit along the curve with arrow, 
the form of the effective potential as a function of 
$\sigma$ changes from a double well to a single well at the edge, HI, of 
the $\sigma$-boundary.  
Hence the value of $\sigma$ drops to a small value as the line moves past the 
edge, signaling a restoration of chiral symmetry at finite temperature. 

We remind that the phase transition along the edge HI is of second order 
since it represents the termination point of a first-order transition along 
the $\sigma$-boundary. Thus the finite-temperature chiral phase transition is 
of second order in the continuum limit.
 
\subsection{$T=0$ and $\mu\neq 0$}

In Fig.~1(c) we show the phase structure at $\mu\neq 0$ with $T=0$.
The effect of finite chemical potential is to create a new phase in the 
parity-symmetric region in between those of $\sigma<0$ and $\sigma>0$.
The new phase is characterized by a small value $\sigma\approx 0$, 
therefore chiral symmetry is effectively restored in it. 
The two phase boundaries JKQP and LMQP are both of first order. 
Hence, when the curve with arrow defining the continuum limit penetrates
the boundary surface, as illustrated in Fig.~1(c), chiral symmetry 
becomes restored through a first-order phase transition. 

\subsection{$T\ne 0$ and $\mu\ne 0$}

\begin{figure}[t]
\centerline{\epsfxsize=6cm \epsfbox{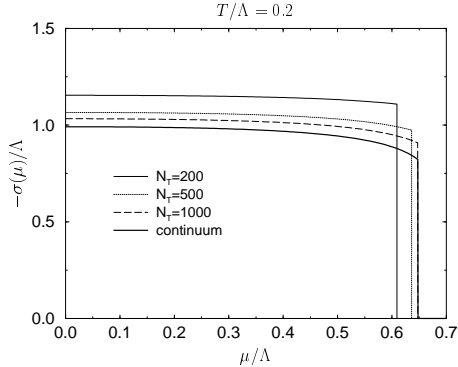}}
\vspace{-1cm} 
\caption{$\sigma/\Lambda$ as a function of $\mu/\Lambda$ at  
at $T/\Lambda=0.2$ for various temporal size $N_T$ together with 
the continuum result.}
\vspace{-0.5cm}
\end{figure}

\begin{figure}
\centerline{\epsfxsize=6cm \epsfbox{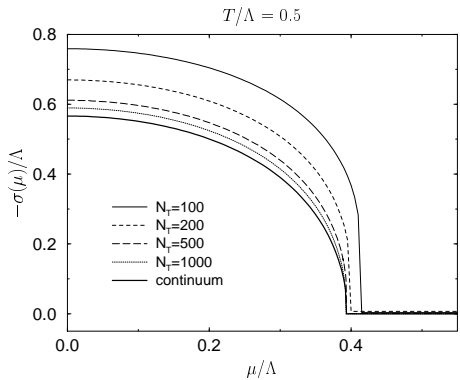}} 
\vspace{-1cm}
\caption{Same as Fig.4 for $T/\Lambda=0.5$. }
\vspace{-0.5cm}
\end{figure}

The phase diagram and order of chiral phase transition for general values 
of temperature and chemical potential are determined by a combination 
of the finite-temperature and finite-density effects discussed above. 
At relatively low temperatures, the situation as a function of $\mu$ is
similar to that at $T=0$.  Hence a first-order chiral transition occurs.
On the other hand, at high temperatures, the behavior of phase structure
is the same type as that at $\mu=0$, leading to a second-order phase 
transition. 

Let us define a $\Lambda$-parameter by
$\Lambda=c\cdot e^{-\pi/g_\sigma^2}/a$ ($ c=0.5716\cdots$) with which 
$\sigma/\Lambda=1$ at $T=\mu=0$ in the continuum limit.
In Fig.~2 and Fig.~3, we plot $\sigma/\Lambda$ as a function of 
$\mu/\Lambda$ for a set of values of $N_T$ at $T/\Lambda=0.2$ and $0.5$. 
We observe a clear first-order transition for the low-temperature case
(Fig.~2), which changes to a second-order transition at high temperatures
(Fig.~3).  Furthermore, the curves for finite $N_T$ 
converge to the continuum result (thick 
solid line) at $N_T\to\infty$.  
These results show how the lattice phase structure leads to the phase diagram 
of the continuum theory with a tricritical point\cite{Wolff}. 

\section{SUMMARY}

The phase structure of QCD in the $(T, \mu)$ plane is still largely 
unknown, and various interesting possibilities have recently been 
discussed\cite{Alford}.  
Lattice study of these possibilities with the Wilson quark action 
has to disentangle effects of parity-broken 
phase from physical ones. Our results should be helpful in the effort in 
this direction.

\vspace*{6mm}
This work is supported in part by Grants-in-Aid of the Ministry of 
Education (Nos. 2375 and 10640246), and by the JSPS Research for Future
Program.  TI is a JSPS Research Fellow.

\end{document}